\documentclass[aps,amsmath,amssymb, pre, twocolumn]{revtex4}
\usepackage{amssymb, amsmath}
\usepackage{bm}
\usepackage{graphicx}
\usepackage{color}
\usepackage{amsfonts}
\usepackage{latexsym}
\usepackage{subfigure}
\usepackage{wrapfig}

\newcommand{\hh }[1]{ \hat{\bm{#1}} }
\newcommand{\m }[1]{ \mathbf{#1} }

\newcommand{\ha }[1]{ \hat{#1} }

\def\eq#1{{Eq.(\ref{#1})}}   
 \def\fig#1{{Fig.\ref{#1}}}

\usepackage{hyperref}

\begin{document}

\title{
Cell motility, 
synchronization, and cell
 traction  orientational order    
 }

\author{M. Leoni  }

\affiliation{  
Physics Department, Syracuse University, Syracuse, NY 13244, USA.}

\begin{abstract}
Suspensions of swimming micro-organisms
provide examples of coordinated  active   dynamics.
 That  has
  stimulated  the study  of a   phenomenological theory combining 
 synchronization  and  polar order  in  active matter  \cite{LLPRL2014}.
 Here, we consider another example inspired 
  by the  traction forces of migrating cells.  
  The  novelty, in this  case, is  the global 
   force-free nature
    of the traction force  field. Such
         a  constraint is absent in the case where  
       the vector field describes 
         swimming speeds in
    micro-organisms suspensions.
     Cell traction  is characterized
by means of a complex tensor quantity,
  that generalizes the nematic orientation
 tensor to incorporate the ability of particles to synchronize, and 
cell motility  depends on this quantity being non-zero.
 We 
 provide a  realization of migrating cell 
which comprises
 an assembly of  dipolar elements exerting traction on a fluid substrate.  
Our model  indicates that spontaneous  transition to the motile state 
  is possible but requires (as in the case of synchronization) non-isochrony of oscillations
  and 
  involves  
subtle synchronization patterns,
 associated to  propagation of waves.
Such results are consistent  with recent experimental work
 relating motility to the synchronization of  actin  oscillators 
 %distributed
%  along
  at the periphery of
  migrating cells.
\end{abstract}

\date{\today}
\maketitle

\subsection{Introduction}

 Motile cells  resemble  microscopic active (i.e. self-driven) droplets. 
 Thanks to this analogy, 
 model cells~\cite{Kruse,Ziebert,Tjhung,ruprecht:2015,bergert:2015}
   which rely on soft active matter theory~\cite{Marchetti-Rev-Mod-Phys,active-gels} 
 have deepened  our understanding of  cell motility.
 So far, however,
  most of these theoretical models  were  focused 
 on paradigmatic cases, like that of
fish-keratocytes~\cite{Mogilner-rev}, where cells form lamellipodia.  
 The motion of such cells involves
regular  flows of actin,    motor contractility \cite{kozlov:2007,rubinstein:2009}, 
     and the interplay with adhesion which
makes it possible  to picture, in theoretical models, adhesion
and 
force-generation  as  time-independent processes. 
 Unsteady oscillatory motions, however,
  are observed in 
other types of   migrating cells, like neurons~\cite{Jiang} or
 amoeba~\cite{Ehrengruber, Tanimoto14, Weiner16, Negrete}.
Analogous behavior is observed also when ameboid cells
 and neutrophils  are suspended in fluids \cite{Barry}.
In adherent  cells~\cite{Burnette}
  forces and adhesion have to be coordinated to enable cell  motion \cite{Giannone} and 
 molecular  determinants   for the control of an excitable, oscillatory,
  system have  been recently identified 
in fibroblasts \cite{Scita18}.

One promising strategy to account for the
spontaneous transition \cite{Spontaneous} 
from  non-motile behavior to  directed migration  
 is based on the following hypothesis:
  coordinated unsteady spatiotemporal
   patterns observed in migrating cells are self-organisation phenomena \cite{self-organis-rev}
    involving feedbacks \cite{Allard} and bio-mechanical interactions \cite{Giannone,Allard} 
    among different sub-cellular structures. 
 The lack of  evidence of any central cellular unit regulating 
 motility and  chemotaxis \cite{Giannone, Allard, self-organis-rev} supports this view.
Synchronization among the various cell parts  is one possibility that 
was invoked in some works studying cell motility \cite{Grzyb,LadouxSYNC}. 
More recently, synchronization of sub-cellular actin oscillators
 was shown to be crucial in controlling the motility of  ameboid  cells \cite{Weiner16}.

 Cell traction forces are essential for
  cell motility, \cite{Trepat}, 
 and play an important role   for cancer research
 where  tumor cells typically exert stronger tractions than control, non-cancerous, cells  \cite{traction-cancer, fabry-3D}. 
The  distribution of cellular  forces  can be conveniently
 characterized using multipolar analysis~\cite{Lau09, Tanimoto14}.
In the absence of externally imposed forces, 
migrating cells are force-free bodies 
 and the monopole term vanishes~\cite{Tanimoto14}.
Dipolar forces have been observed in adherent cells both at the scale of the entire
 cell~\cite{De,Schwarz} and at subcellular scales~\cite{Ghassemi}.
A dipolar force distribution  is  however  front-back symmetric. Therefore,
the dipole term alone is not enough to  justify a preferred direction for
 persistent cell motion, which might be due  to the existence of a quadrupolar  
  force distribution~\cite{LS15, LS17}. 
 
This theoretical  consideration has been shown to hold true by 
 experimental work~\cite{Tanimoto14}  which
  characterized motility of {\it D. discoideum} 
   via traction force microscopy. 
 These authors  confirmed that both   dipole and  quadrupole terms
 provide robust tools for describing migrating cells -- 
 showing that while the axis along which the motion occurs is 
correlated  with the orientation of the dipolar forces, the 
  direction of motion  is rather determined by  the 
  quadrupolar term along that  axis.

{

The study of   synchronization  coupled to orientational dynamics (e.g. dipole and quadrupole terms)
  seems thus  relevant for cell motility.
So far, to our knowledge, only a few theoretical studies
 focused on the interplay of orientational order and synchronization. This was done  in different contexts: 
studying the interplay of polar order and synchronization in soft active fluids \cite{LLPRL2014};  
considering swarmalators \cite{swarmalator}, which  combine the ability to swarm and 
to synchronize; investigating collections of self-propelling particles which can synchronize
 \cite{Pagonabarraga,Pagonabarraga2}.
Experimentally,
 a work on   bacterial suspensions
 interpreted   collective oscillations 
     as frequency synchronization  of bacterial orientational dynamics \cite{synchro-bacteria}.

The implications of synchronization for
 cell motility have been  examined  in~\cite{LS15}, 
considering the
interplay of intracellular and extracellular mechanical interactions
for a model cell with   dipolar forces~\cite{LS15}, already {\it aligned},  in contact with a  fluid substrate. 
Such microscopic dipolar forces yield   dipolar and quadrupolar terms at the cell scale. 
In the same spirit, another recent work~\cite{LS17} investigated the role of stochastic adhesion in
 model cells
with oscillatory  
force-distributions 
 already aligned and synchronized.

Here we 
 pursue the study of synchronization and cell motility considering also  orientational order. 
We adopt an approach 
 which generalizes the use of  conserved and broken symmetry variables  \cite{Chaikin} 
 to characterize the active non-equilibrium dynamics   \cite{Marchetti-Rev-Mod-Phys} at scales larger than individual microscopic elements. 
For motile cells, such elements might be  acto-myosin assemblies forming micron-sized sarcomeres \cite{Wolfenson}.

The one particle concentration  
\begin{align}
& c(\m x, \hh u, \phi, t) = 
 \nonumber \\
&
 \langle \frac{1}{N} \sum_N \delta(\m x- \m x_n(t)) \delta(\hh u -\hh u_n(t))  \delta(\phi -\phi_n(t))\rangle
\label{eq:c}
\end{align}
  is associated to the probability  of finding an active element with position $\m x$, orientation $\hh u$, phase $\phi$ at time $t$
 given  the microscopic 
 dynamics 
of  active elements with positions 
$\m x_n(t)$, orientations $\hh u_n(t)$ and  phases  $\phi_n(t)$,
for $n = 1, \ldots N$.
The large scale behavior of active matter systems,  \cite{Marchetti-Rev-Mod-Phys},
 can be characterized using  
vector and tensor quantities that are 
related to the vector character (via $\hh u$) of the microscopic elements.
 In two dimensions, 
density, $\rho$, polarization, $\m p$, and nematic orientation tensor, $\mathbb{S}$,  
are written as moments of the concentration as follows
 \begin{align}
 & \rho(\m x, t)  = 
 \int d\hh u \int d\phi   
  c(\m x, \hh u, \phi, t ) 
  \label{eq:rho} \\
 & \m p(\m x, t)  = 
\int d\hh u \int d\phi   
 \hh u
  c(\m x, \hh u, \phi, t ) 
  \nonumber \\
    & \mathbb{S}(\m x, t)  = 
  \int d\hh u  \int d\phi   
[
 \hh u \otimes \hh u - \frac{\mathbb{I}}{2}
 ]
  c(\m x, \hh u, \phi, t ) 
  \nonumber 
   \end{align}
  Here $\rho$ is a conserved quantity while
  $ \m p $ and $ \mathbb{S}$ are broken symmetry variables:  order parameters describing the degree of polar order or
  of nematic order in the system.

In systems displaying   synchronization,
 however, the phase $\phi$ of each  element provides an additional degree of freedom which is not  
 taken into account in the above system of moments. 
 To incorporate the effect of the phase,
 another series of moments 
 generalizes the above construction
 as follows
 \begin{align}
 & \Phi(\m x, t)  = 
 \int d\hh u  \int d\phi  
 e^{i\phi} 
  c(\m x, \hh u, \phi, t ) 
 \label{eq:mom-new}   \\ 
   &
    \bm{\Pi}(\m x, t)  = 
 \int d\hh u  \int d\phi 
  e^{i\phi}    
 \hh u 
  c(\m x, \hh u, \phi, t )
     \nonumber \\
     & \Sigma(\m x, t)  = 
  \int d\hh u  \int d\phi     e^{i\phi}  
[
 \hh u \otimes \hh u - \frac{\mathbb{I}}{2}
 ]
  c(\m x, \hh u, \phi, t ). 
  \nonumber 
  \end{align}
The first two  terms in \eq{eq:mom-new}  were already obtained in  \cite{LLPRL2014}: 
$\Phi(\m x, t) $, in the spatially homogeneous limit, is 
 the Kuramoto's order parameter \citep{Ritort}  and 
  $ \bm{\Pi}(\m x, t)$ describes the combined effect of
vectorial symmetry and synchronization. 
The last  moment,  $\Sigma(\m x, t)$,  
  generalizes   the nematic orientation tensor $\mathbb{S}$  in presence of 
   synchronization
   just like  $ \bm{\Pi}(\m x, t)$ generalizes $\m p(\m x, t)$.

There is however a crucial difference between  
 $ \bm{\Pi}(\m x, t)$ and
 $\Sigma(\m x, t)$: 
 the former can be employed to describe 
 collective dynamics that has no global constraints, like in the case of
   suspensions of micro-swimmers
where $\hh u$ in  \eq{eq:mom-new}  is associated to the orientation (or speed) of the  individual swimmer; 
the latter is suitable for  describing global constraints, like in the case of
  traction forces exerted by migrating cells
 where $\hh u$ represents the director of  the cell traction  vector field which  is constrained to be  force-free. 
We note that  $\Sigma(\m x, t)$ could be exploited  also in other  contexts,
 like the dynamics of  
  chromatin inside nuclei \cite{chromatin} where specific enzymes,  intervening during 
   chromatin remodeling,   exert   local active dipolar forces \cite{Grosberg}.

\paragraph{No global constraints}
Using \eq{eq:c},  
 one can write $\bm{\Pi}(\m x, t)  $ as    
 \begin{equation}
\bm{\Pi}(\m x, t)  
=\frac{1}{N} 
 \langle 
\sum_n \hh u_n e^{i \phi_n} \delta(\m x - \m x_n)
\rangle.
\label{eq:Pi0}
 \end{equation} 
 \eq{eq:Pi0}
 shows that  
in the special case of  spatially homogeneous polar state, where
all the directors point in the same direction, say 
 $\hh u_n \equiv \hh U$ $\forall n$, then
$\m p \equiv  \hh U $ and the complex vector order parameter  $  \bm{\Pi}(\m x, t) $
 is just  the product of the two order parameters, one describing polar order and the other synchronization, 
as  $  \bm{\Pi} \equiv \m p \Phi $. 
The same holds true for a  synchronized state, 
where  $\phi_n \equiv \varphi$,  $\forall n$, and again   $  \bm{\Pi}(\m x, t) $ is
 the product of
$\m p$ and $\Phi$. However
there are states where  
  $  \bm{\Pi}(\m x, t) $ describes non-trivial configurations. 
  For example, one can construct 
  states where both $\m p = 0$ and $\Phi =0 $ and yet
  $  \bm{\Pi}(\m x, t) \neq 0$  as it was pointed out in \cite{LLPRL2014}.

  \paragraph{Force-free constraint}
A migrating cell is a force-free system, meaning that the forces exerted by the cell add up to zero,
 as experimentally 
verified  e.g. in \cite{Tanimoto14}.  
 As it will be shown below, this constraint 
  lead us to consider    $\Sigma(\m x, t)$ for describing cell motility.
Using \eq{eq:c},  
 one can    
  relate $\Sigma(\m x, t)$  to the microscopic variables as 
 \begin{equation}
 \Sigma(\m x, t)  
=\frac{1}{N} 
 \langle 
\sum_n (\frac{\mathbb{I}}{2} -  \hh u_n \hh u_n ) e^{i \phi_n} \delta(\m x - \m x_n)
\rangle. 
\label{eq:Sigma1}
 \end{equation}
 In earlier phenomenological studies on active systems which do not include the internal cyclic dynamics, \cite{Rama},
 it  was already noted that the  force density of   force-free active particles 
 has   nematic symmetry. 
\eq{eq:Sigma1}   recovers that result    in the case where   the cyclic elements
 have all the same phases, $\phi_n = \phi_0(t) $,
  $\forall n$. Then, from \eq{eq:Sigma1} 
 $$ 
\Sigma(\m x, t)  
= \Phi 
\mathbb{S} 
 $$  
 where $ \Phi = e^{i \phi_0(t)}   $.
  In particular, we can recover the regime discussed in \cite{Rama}, and in other following works on active matter
  \cite{Marchetti-Rev-Mod-Phys},
   by posing $\phi_0(t)  = const$. This means that  there is no cyclic dynamics and
  $\Phi  $ is just a normalization factor in front of the nematic orientation tensor order parameter, $\mathbb{S}$.

 The article is organized as follows:
 in Section \ref{sec:1} we introduce a microscopic model of the cell
coupled to the substrate
in 
 Section \ref{sec:2}, relying on that model, 
we study  the cell  migration speed. We use analytical methods to show how the
 speed relates to
the broken symmetry variables \eq{eq:mom-new} and 
investigate the speed numerically for simple configurations.
Finally, in 
 Section \ref{sec:3} we present the
dynamic equations for  \eq{eq:mom-new} discussing the links with  motility.

\subsection{Model cell  on a fluid substrate}
\label{sec:1}
In this section we provide a concrete realization
 where, 
by construction,
the vector field associated to the active traction forces exerted by
 a cell
  satisfies the force-free constraint:
 a set of dipolar units
distributed in a quasi 2D region of space describing the cell boundary, see \fig{fig:scheme}.

\begin{figure}[h!]
 \includegraphics[scale=0.3]{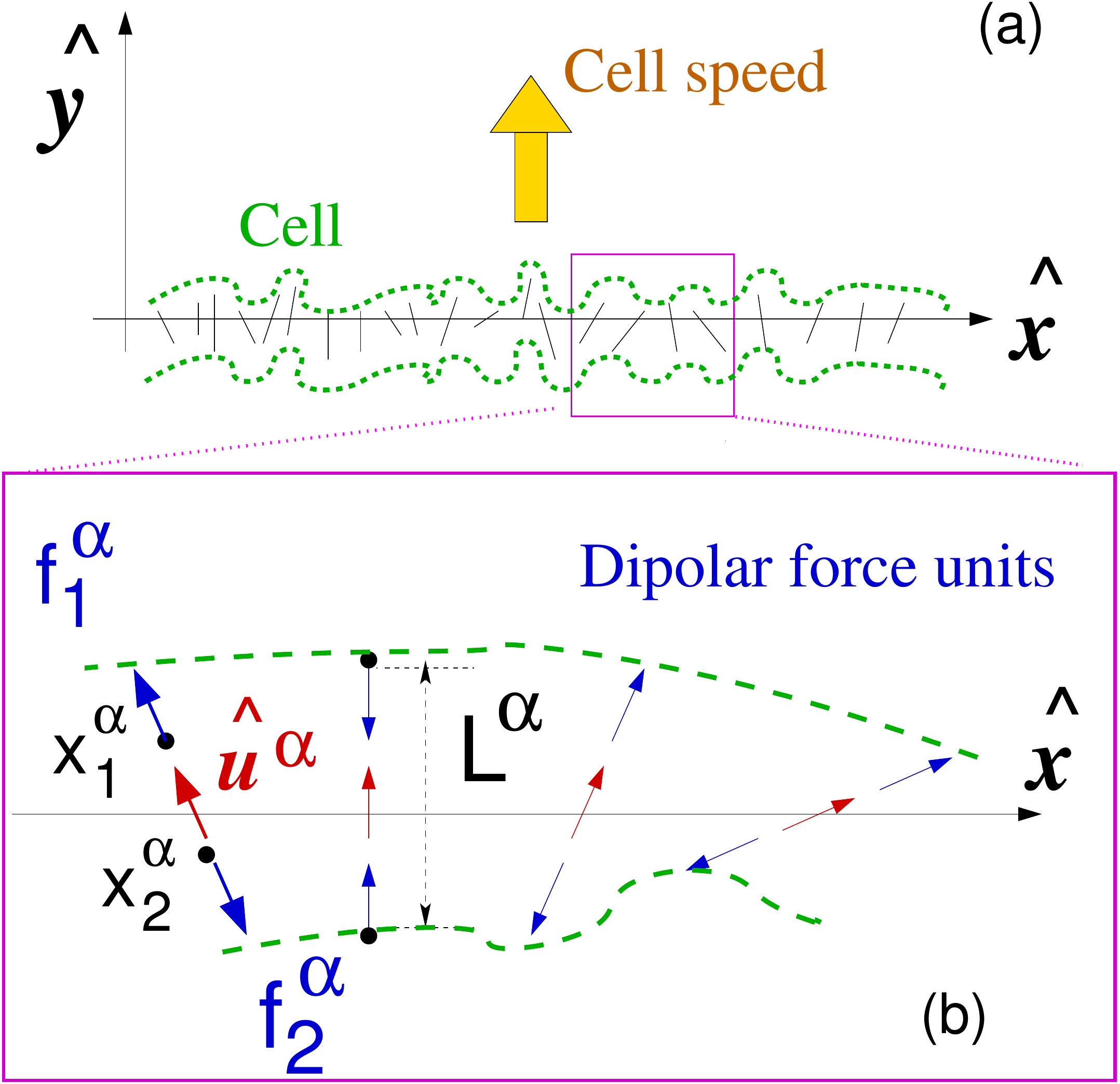}
\caption{{\it Quasi}-2D model of a motile cell. 
 (a) Model cell depicted as a collection of dipolar traction force units  which are distributed  along the x-axis.
   (b)  Zooming the   details of (a).   Unit vectors $\hh u^\alpha$ (labeled with $\alpha = 1, \ldots M$) determine the traction unit which comprises two equal and opposite forces
   $\m f^\alpha_{n} = (-)^{n+1} F^\alpha \hh u^\alpha $ at distance $L^\alpha$.
    Note that, despite the fact that the centers of the unit vectors $\hh u^\alpha$ are along one dimension, the directions 
     $\hh u^\alpha$  can span the entire plane. 
    }
    \label{fig:scheme}
\end{figure}

 Experimental measurements %of sub-cellular force generating units 
 indicate the existence of sub-cellular contractile units
in adherent cells, which can be 
detected and measured using micro-pillars, see e.g. \cite{Ghassemi, Wolfenson}. The size of these contractile
 units is of the order of a few microns  and it has been suggested that they are acto-myosin micro-sarcomeres \cite{Wolfenson}.
Similar sub-cellular traction units are
seen also in migrating cells \cite{Jiang}  
and can be visualized
 by means of 
fluorescence \cite{draft-Riveline}. 
The intensity of the fluorescent signal might
be a measure 
of the intensity of the active 
force exerted by a given cell, useful for  
comparisons   with theoretical models.
 
 \subsubsection{Traction force unit}
The traction force unit, labeled with $\alpha$, is a dipolar element
  made of two particles,  in turn labeled with  
$n=1,2$.
Each particle is driven by an active force 
$
\m f^\alpha_n$. 
We  write  $\m f^\alpha_{n} =  (-)^{n+1} \m F^\alpha $ 
and 
$ \m F^\alpha =  F^\alpha \hh u^\alpha   $. For now, we pose
 $ F^\alpha = f_0 \cos \phi_\alpha  $.
As by construction  $ \hh u^\alpha_1  = -  \hh u^\alpha_2 $, see \fig{fig:scheme},
 the total force of the dipolar unit is zero,  
$\m F^\alpha =  \m f^\alpha_1 + \m f^\alpha_2 = f_0  \cos \phi_\alpha ( \hh u^\alpha_1 +  \hh u^\alpha_2) = 0 $. 
The same remains true when summing all the forces at the level of the entire cell.
The cell is thus force-free.

 \subsubsection{Microscopic force-density}
The  force density 
$\bm{ \mathcal{F} }(\m x)$  on the surrounding medium is obtained considering
 the microscopic density function, given by the Dirac's delta `function',
\begin{equation}
\bm{ \mathcal{F} }(\m x) = f_0  \sum_\alpha   \cos \phi_\alpha \sum_n  \hh u^\alpha_n \delta(\m x - \m x^\alpha_n ) .
\end{equation}
This force distribution is what enters at r.h.s. of the  continuum equation describing the 
medium, see e.g. \cite{Rama, Marchetti-Rev-Mod-Phys} and \eq{eq:stokes} below.

To gain insight, we expand the delta function \cite{Rama},  writing 
$\m x^\alpha_n =  \m c^\alpha +  \frac{L^\alpha}{2}    \hh u^\alpha_n$
where $ \m c^\alpha $  is the coordinate of the center  of the traction unit $\alpha$.
Note that the first term of the expansion generates a contribution
 of the form
$ 
\sum_\alpha   \cos \phi_\alpha \delta(\m x - \m c^\alpha ) \sum_n  \hh u^\alpha_n  = 0 
$
which 
vanishes  due to the  force-free condition.
This equation can be translated 
in terms of the complex vector order parameter,
$\bm{\Pi} $, compactly as 
$Re[ \bm{\Pi}^0 ] = 0 $
where the over-script $0$ denotes a spatially homogeneous value.
Hence,  for spatially homogeneous states,
 $\bm{\Pi}^0$  can be associated to the monopolar term of the traction forces.

The next term in the expansion is obtained by noting that
$  ( \hh u^\alpha_1  \hh u^\alpha_1 +  \hh u^\alpha_2  \hh u^\alpha_2) = 2 \hh u^\alpha_1  \hh u^\alpha_1 $.
We further simplify our notation by posing $\hh u^\alpha := \hh u^\alpha_1$.
As a result we are left with
$\bm{ \mathcal{F} }(\m x) \approx  f_0
 \sum_\alpha   L^\alpha   \cos \phi_\alpha   \hh u^\alpha    ( \ha u^\alpha)_k \nabla_k \delta(\m x - \m c^\alpha ) .
$
If $L^\alpha $ performs  small oscillations around $l_0$ we obtain
\begin{equation}
 \langle  \mathcal{F}_i(\m x) \rangle \propto  f_0 l_0  Re[ \nabla_k \Sigma_{ik} ( \m x) ]
\end{equation}
where $Re[...]$ indicates the real part of the term in parenthesis.
As anticipated,
 this generalizes
 the results obtained in \cite{Rama} to the case of 
particles which possess internal cyclic dynamics.
Here $ f_0 l_0$ quantifies the magnitude of the microscopic dipole
while  $\Sigma_{ik}$ incorporates both orientations and   phases.

  \subsubsection{Cell coupled with the extra-cellular medium}
  In order  
  to migrate, a cell needs 
  to exert forces on its surroundings.
  In vivo,   cells are surrounded by an
  extra-cellular matrix, a polymer gel,
  that can be reproduced also in vitro \cite{Muller,Yamada3D}.
Here we discuss 
   the viscoelastic dynamics of the gel in the long time limit \cite{Kruse} 
 where the gel is pictured as a viscous fluid. 
 
 Although the fluid model is used here as a simple example of 
   extra-cellular medium, this approach is not 
 unrealistic:  a fluid description of an elastic substrate, directly compared with experiments, was already
  given in
  \cite{Peterson}.
Furthermore,
 in elastic substrates 
  the force-distribution
determines substrate deformations.  
 Deformations alone
  would suffice for describing 
the dynamics of adherent cells
 but
 they are not enough for migration.
 In fact,
  for   migrating cells 
   one has to include also the adhesion
    dynamics  
to  describe how 
a cell unbinds and subsequently rebinds at different positions on the substrate. 
A model of transient, stochastic, adhesion coupled to cell forces and mechanics
 was proposed in \cite{LS17}. 
That framework can  be generalized  to   a 
soft, deformable, substrate \cite{LS-unpub} but is not considered here.

The equation describing an incompressible  viscous fluid 
of viscosity $\eta$ and velocity $\m v$
 in the absence of inertia is
 \begin{equation}
  \eta \nabla^2 \m v - \bm{\nabla} p = -\bm{ \mathcal{F} }(\m x) ; \qquad  \bm{\nabla} \cdot \m v = 0.
  \label{eq:stokes}
 \end{equation}
 This equation determines the 
 velocity $\m v(\m x)$ generated in the surrounding viscous  medium, given the force distribution
 $\bm{ \mathcal{F} }(\m x)$. In turn, one can express 
 $\m v$ as an integral of $\bm{ \mathcal{F} }$ using Green's functions formalism \cite{Doi}. The component $v_i$ of the 
 velocity is given by
\begin{equation}
  v_i (\m x) = \int d\m y  H_{ij}(\m x- \m y)    \mathcal{F}_j(\m y) 
\label{eq:v}
\end{equation}
(where we use Einstein's summation convention on repeated indexes  for the component $j$).
 This allows us to connect the
  theory
 to the experimental measurements on the velocity of the substrate. 
For a semi-infinite substrate, with flat surface,  $H_{ij}(\m r) \sim \frac{\delta_{ij}}{2\pi \eta r} $, see \cite{LS15}.

The dynamic equation of a cell in contact with the medium is obtained from
force balance ($2^{nd}$ Newton's law). With the choice of fluid medium, and  neglecting inertia, 
the equation is
\begin{equation}
 0 \approx -\gamma ( \dot{\m x}^{\alpha}_n -  \m v(\m x^{\alpha}_n) ) + \m f^\alpha_n 
\label{eq:force-balance}
\end{equation}
where $n = 1 \ldots N$ and $\alpha = 1, \ldots M$ label different particles representing cellular adhesion sites \cite{LS15, LS17}. 
 $\gamma$ here is a friction coefficient (for a semi-infinite substrate is related to the Stokes' drag, $\gamma = 3 \pi \eta a$ where $ a $ is the radius of the particle \cite{LS15}) and 
$\m v(\m x^{\alpha}_n) $ is the flow generated at position $\m x^{\alpha}_n $ due to all the remaining particles,
which can be computed using \eq{eq:v}.

At this point  there is still one thing to specify, namely 
the dynamics of the active forces (or equivalently, the dynamics of the phases $\phi_\alpha$ as forces depend on these variables).
We will address this  in the following.

\subsubsection{Dynamics of the force generators}

The traction force
 elements are subjected to active forces and coupled with the substrate (here  a fluid).
 The term ``active'' here means that forces 
  vary in time. For cyclical variations,
   the time-dependence can be 
 specified using a phase variable. Once this is set,
from \eq{eq:force-balance}  we  derive the dynamics of other variables 
such as orientations, amplitudes and phases of oscillations.
The dynamic equations for such variables are  
 needed to derive  the equations for the moments, \eq{eq:mom-new},
describing the large scale, coarse-grained, dynamics.

\paragraph{Orientational dynamics}
By construction, in our model, 
there is no torque on the traction units. Hence the traction units cannot 
 rotate when isolated but they can rotate thanks to mechanical interactions 
 with other  units. 
The angular speed  ${\omega}^\alpha $ of such rotations 
 can be computed from $\dot{ \hh u }^\alpha $ via
 $ \dot{\m L}^\alpha = d(  L^\alpha  \hh u^\alpha)/dt$, see \fig{fig:scheme}(b), 
as 
${\omega}^\alpha  =
\frac{ \hh u^\alpha }{L^\alpha} \wedge  (\dot{\m x}^\alpha_1- \dot{\m x}^\alpha_2)$. 
Using \eq{eq:force-balance} 
we obtain
\begin{align}
&
\omega^\alpha_a  \approx  
 \epsilon_{abi}  \ha u^\alpha_b  \sum_{\beta \neq \alpha}     \frac{\mathcal{O}^{\alpha \beta }_i }{L^\alpha }.
 \label{eq:omega}
\end{align}
Here  $ \mathcal{O}^{\alpha \beta}_i$ follows  from
\eq{eq:v}. 
Its detailed expression is reported in the appendix \ref{sec:app-orient}.
Note that the orientational dynamics contributes to  the dynamic equation for   the moments $\m P, \mathbb{S},
 \bm{\Pi}, \Sigma$ \eqref{eq:mom-new}. 
However for  $\Sigma$, which is one of the relevant quantities, the contribution of   $\omega^\alpha$ is sub-dominant compared to that of the phase dynamics, as discussed below.  
 Moreover,
  in the study of the cell speed done below, in \ref{sec:cell-speed},
    for simplicity
  we consider  configurations where the orientations
   $\hh u^\alpha$ lie along a given direction and we neglect their 
  dynamics. 

To derive the equation for the amplitude and phase describing oscillatory dynamics we
first need to obtain the  equations regulating the deformation and the forces of the
traction units. 
 \paragraph{Deformation dynamics of a traction unit}
The  
deformation
 dynamics of each dipolar traction unit  follows from the definition of $\m L^\alpha = {\m x}^\alpha_1- {\m x}^\alpha_2$,   see \fig{fig:scheme}(b),  as
$\dot{ L}^\alpha  =
\hh u^\alpha \cdot (\dot{\m x}^\alpha_1- \dot{\m x}^\alpha_2)$.
Hence, the dynamic equation for the  internal deformation of the element $\alpha$  
is obtained from \eq{eq:force-balance} as 
\begin{align}
&
\dot{ L}^\alpha  
= 2  F^\alpha/\gamma  +
 \sum_{\beta \neq \alpha}    \hh u^\alpha \cdot \bm{ \mathcal{I}}^{\alpha \beta}
 \label{eq:Ldot}
\end{align}
Here, $ \bm{ \mathcal{I}}^{\alpha \beta}    $ 
results
 from  
 \eq{eq:v}. 
Its detailed expression is reported in the appendix \ref{sec:app-def}.
To leading order,
$\dot{ L}^\alpha  \approx 
2  F^\alpha /  \gamma
$
 has  solutions
$F^\alpha = - f_0 \sin(\omega_0 t + \phi_\alpha ) $
and 
$L^\alpha = 2 \frac{ f_0}{\gamma \omega_0}  \cos(\omega_0 t + \phi_\alpha )$.  
Thus, neglecting mechanical interactions  $ \mathcal{I}^{\alpha \beta}_i$ and in the absence of noise,  
the oscillating dipoles will maintain  
their relative phase difference which is controlled solely
 by the initial conditions. 
More interestingly in presence of 
 interactions  the phases can vary.
To describe how   the forces, and hence the phases, evolve we use a  generic model
 of self-sustained oscillating forces  introduced in \cite{LLPRE2012}.

\paragraph{Force  dynamics of a traction unit}
The evolution of the forces, following \cite{LLPRE2012},  is described by 
\begin{equation}
 \dot{F}^\alpha = -\mathcal{K}  d_\alpha + \mathcal{M} 
F^\alpha (1-  \mathcal{S}  d^{2}_\alpha )/\gamma  + \mathcal{A}  d^{3}_\alpha  
\label{eq:Fdot}
\end{equation}
where 
$L^\alpha = l_0 + d_\alpha $. As for the other parameters in \eq{eq:Fdot}, 
$\mathcal{K} > 0 $ determines the frequency of oscillations, 
 $\mathcal{M} > 0 $ yields self-sustained oscillations,
 and $\mathcal{A}$, which can be either positive or negative,
  determines  the non-isochrony of the oscillations.
  $ \mathcal{S}$, associated to the saturation of oscillation amplitude, will be set equal to 1 in the following.
 As explained   in the appendix \ref{sec:app-phase-ampl}, we can map \eq{eq:Fdot}
and \eq{eq:Ldot}  onto
  equations for the phase 
 for the amplitude.

\paragraph{Amplitude and phase dynamics}
\label{sec:ampl-phase}
Complex amplitudes are related to $L^\alpha$ via 
$L^\alpha \sim l_0 +  [ A_\alpha e^{i \omega t}  + c.c. ]/2$.  
Similarly for the force, we pose
$F^\alpha \sim i  \gamma \omega_0 [ A_\alpha e^{i \omega t}  - c.c. ]/4$.  
In turn,
 the complex amplitude is related to real amplitude and phase  $ R_\alpha$, $  \phi_\alpha $ via
 $A_\alpha = R_\alpha e^{i \phi_\alpha }$.
  Using these relations,
    we obtain an equation for the real amplitude
   \begin{align}
 &  \dot{R}_\alpha \sim 
(\lambda  - \beta R^2_\alpha  ) R_\alpha
+  l^2_0    \ha u^\alpha_i \ha u^\beta_j (  \ha u^\alpha_k \ha u^\beta_l +  \ha u^\alpha_l \ha u^\beta_k  ) \times
\label{eq:Rdot}  \\
&
\frac{R_\beta \gamma  \omega_0 }{4} \sin( \phi_\beta  - \phi_\alpha ) T_{ijkl}(x_\alpha - x_\beta)
 \nonumber 
   \end{align}
with
$ \lambda = \frac{\mathcal{M}}{\gamma} $;
$\beta = \frac{\mathcal{M}}{4 \gamma} $ and  $T_{ijkl}(x_\alpha - x_\beta) = \nabla_k \nabla_l H_{ij}( x_\beta  - x_\alpha )$.

We consider  interactions acting  as a small perturbation to 
the 
non-interacting dynamics.  In this regime,  
 \eq{eq:Rdot}
describes deviations from a  fixed point dynamics.
The fixed
point is
 the limit cycle  ${R}_\alpha \sim R_0 $
with  
$ R_0 
=  \sqrt{\lambda /\beta} .
$
To study 
how these perturbations evolve,
 we
write
$ {R}^\alpha \sim  R_0 + \delta  {R}^\alpha $
and substitute this expression in \eq{eq:Rdot} keeping
terms up to order $\delta$.

The interpretation of the dynamics of $\delta R_\alpha$ has been given elsewhere for 
oscillators moving along one dimensions \cite{LLPRE2012}. 
We summarize the main points:
1) due to interactions, the oscillators' trajectories  in the phase space
 move away from the limit cycle. This is described by terms $\delta R_\alpha$; 
2)   the limit cycle is a stable fixed point:
  deviations  $\delta R_\alpha$  relax 
 to the limit cycle and behave as  dumped fluctuations.
The result is essentially the same 
here  with the main change 
  due to the presence of the directors $\ha u^\alpha_k $ 
  (as motions are no longer in one dimension).

Setting $ \dot{ \delta R}_\alpha = 0 $
we can eliminate  $ \delta R_\alpha $
 in favor of the phase, see appendix \ref{sec:app-phase-ampl},
  obtaining the equation for the phase dynamics
 \begin{align}
 &  \dot{\phi}_\alpha \sim 
- \Delta - \chi R^2_0 
 - \frac{ l^2_0 \omega_0 \gamma }{4 } 
 \ha u^\alpha_i \ha u^\beta_j (  \ha u^\alpha_k \ha u^\beta_l +  \ha u^\alpha_l \ha u^\beta_k  )
   \times
\nonumber
 \\
&
[  \cos( \phi_\beta  - \phi_\alpha ) + 
\frac{ \chi    }{\beta  }    
 \sin( \phi_\beta  - \phi_\alpha )
]T_{ijkl}(x_\alpha -x_\beta) 
 \label{eq:Phidot2} 
   \end{align}
where $\Delta = - \frac{\omega_0}{2}  +  \frac{\mathcal{K}}{\gamma \omega_0}$ and 
$\chi = 
\frac{3 \mathcal{A}}{4 \gamma \omega_0}.
$
We note that,  in particular,  the sinusoidal term of \eq{eq:Phidot2} is 
responsible for synchronization \cite{LLPRE2012}.

 \subsection{Cell speed}
 \label{sec:2}

 \label{sec:cell-speed}
   From
  the mean  position of the dipolar units
 ${ \m C} = \sum_n \sum_\alpha {\m x}^\alpha_n/M $ 
 (for $\alpha = 1, \ldots M$)
  we obtain
  an important readout for cell motility:
   the cell speed 
\begin{equation}
  \dot{ \m C} =\frac{1}{ M} \sum_n \sum_\alpha \dot{\m x}^\alpha_n 
\label{eq:Cdot}
\end{equation}  
which can be computed using \eq{eq:force-balance}.
The choice of a {\it quasi}-2D model, see \fig{fig:scheme}, simplifies the calculation. In fact,  
we can write 
$\m x^\alpha_n = x_\alpha  \hh x + (-)^{n+1} \frac{L^\alpha}{2} \hh u^\alpha $
and decompose the problem along two independent directions, $\hh x$ and $\hh u^\alpha$.
 Performing an expansion valid at distances 
 larger than $L^\alpha$ (consistent with our long wavelength description) 
 the 
 tensorial part $H_{ij}[ (x_\alpha - x_\beta)  \hh x] $ only depends on the director $\hh x$ while the other variables are defined in the
 whole   $(x,y)$ plane.
We note also that
 the exact form of $H$ does not play a major role for what concerns the purpose of this work.
  In fact,
due to the {\it quasi}-2D nature of the forces, where the centers $x_\alpha$ of the force-traction units lie 
 along the x-axis, see \fig{fig:scheme},
  different choices of $H$ lead to quantitative  but not qualitative 
 differences.

\subsubsection{Analytical study}
\label{sec:cell-speed-anal}
To make progress, we need to expand
the term $H$ describing interactions.
Consistently with the analysis in
 \ref{sec:ampl-phase} we shall assume that dipolar   forces and  deformations depend on time as
$F^\beta = - f_0  \sin(\omega_0 t + \phi_\alpha ) $
and 
$ L^\beta = l_0 + R_0 \cos(\omega_0 t + \phi_\alpha ) $.
The expansion 
 produces 
 terms of the form 
  $F^\beta L^\beta \sim \sin 2 (\omega t + \phi_\beta) $
which average to zero over a period $T = 2\pi /\omega$. 
However,  terms containing  $F^\beta L^\alpha $   give a finite contribution  for $\alpha \neq \beta $.
Using a mean-field approach \cite{LL2010PRL}
the speed  is given by 
\begin{align}
&
\langle
 \dot{ C}_i
 \rangle
\propto 
\frac{1}{24i}
\int d  x_\alpha  \int d  x_\beta
U_{ijklu} (x_\alpha - x_\beta)
 \mathcal{G}_{jklu} [ \Sigma,   \Phi  ] .
\label{eq:Cdot2}
 \end{align}
Here the term  
$U_{ijklu} (x_\alpha - x_\beta) = \nabla_k \nabla_l \nabla_u H_{ij}[ (x_\alpha - x_\beta)  \hh x] $ describes
the dominant contribution of the 
 interactions between different dipolar units. 
Along with the interactions, the speed is determined by the term in parenthesis in \eq{eq:Cdot2}.
The functional  $ \mathcal{G}_{jklu}[ \Sigma, \Phi  ]$ is defined as 
\begin{align}
& 
 \mathcal{G}_{jklu}[ \Sigma, \Phi  ] = 
\{ \Sigma^{\alpha}_{kl} 
(\Sigma^\beta_{uj} 
)^*
- c.c. 
 \nonumber \\
 & + 
 \Phi^{\alpha}
(\Sigma^\beta_{uj})^*  \delta_{kl}/2  - c.c 
\}
 + 
\{ u \leftrightarrow l  \}
\label{eq:G}
\end{align}
where we have used the shorthand  $\Sigma^\beta_{uj} = \Sigma_{uj}( x_\beta)$ and similarly for $\Phi^{\alpha}$. 
  $ \{ u \leftrightarrow l  \}$ indicate a contribution similar to the previous
   expression in parenthesis but with the indexes $u$ and $l$ exchanged.
  
\eq{eq:Cdot2}  and \eq{eq:G}
  confirm the need of a phase difference along the cell.   
Since the speed is controlled by 
  $\Sigma$ and $\Phi$, which in turn
are functions of two microscopic degrees of freedom
$\hh u$ and $\phi$,
 it is instructive to examine the impact of these
variables separately.

  \paragraph{Synchronized and spatially homogeneous states}
  \label{sec:speed-hom}
We begin with examining the case of
 the phases which are identical everywhere.
 This means that
  $\Sigma^\beta_{uj} = \Phi^0  \mathbb{S}^\beta_{uj}$
and 
$$  
 \Sigma^{\alpha}_{kl}
 (\Sigma^\beta_{uj} )^* - c.c.   = |\Phi^0|^2 [ 
\mathbb{S}^\beta_{uj}\mathbb{S}^{\alpha}_{kl} 
-\mathbb{S}^\beta_{uj}\mathbb{S}^{\alpha}_{kl} 
] = 0. 
$$
Likewise, terms of the form
$$  
\Phi^{\alpha}
 ( \Sigma^\beta_{uj} )^*  \delta_{kl} - c.c.   = |\Phi^0|^2 [ 
\mathbb{S}^\beta_{uj}\delta_{kl} 
-\mathbb{S}^\beta_{uj}\delta_{kl} 
] = 0
$$
so the associated speed in \eq{eq:Cdot2} is zero.
The same conclusion is reached in the 
(less general)
 case of 
  spatially homogeneous order
 where,
  in addition to $  \Phi^\beta= \Phi^\alpha =   \Phi^0 $,
   one has  
 also $ \Sigma^\beta_{uj} = \Sigma^\alpha_{uj} =   \Sigma^{(0)}_{uj}$.

\paragraph{Non-synchronized states}
Let us now consider instead the alternative scenario 
 where
the orientations are all constrained (e.g. along $\hh x$ -- see also numerics  below) 
while the phases are  free to vary and may synchronize. We pose 
$ \Sigma^\beta_{uj} =  \mathbb{S}^{0}_{uj}  \Phi^\beta  $
 (with e.g.  $ \mathbb{S}^{0}_{uj}  = ( \frac{\delta_{uj}}{2} - \ha x_u \ha x_j )$).
In this case
\begin{align}
&
 \mathcal{G}_{jklu}  [\Sigma ,   \Phi  ]   = 
 \{
\mathbb{S}^{0}_{uj}
\mathbb{S}^{0}_{kl}
  [ 
\Phi^*(x_\beta)  \Phi(x_\alpha) 
- c.c. 
] 
\nonumber \\
& 
+
\mathbb{S}^{0}_{uj}
 \frac{\delta_{kl}}{2} 
  [ 
  \Phi(x^\alpha) 
-
c.c.
]
\}
  + \{ l \leftrightarrow  u \}
\end{align}
and a non-zero cell speed in this case is associated to inhomogeneities  
of the  order parameter $\Phi$.
In the particular case of just two traction units oriented along the $\hh x$ axis we recover what expected from 
 previous analysis, 
\cite{LS15}, 
  for a cell made up of two units $A$ and $B$
where cell speed scales as 
$ \dot{C} 
\sim 
\nabla^3 H(r) \sin(\phi_A - \phi_B ). 
$

\paragraph{Wave-like perturbations}
 To gain insight, we consider a wave propagating  along $\hh x$ and
 consider 
  the effect of such a perturbation on the speed retaining terms
   up to linear order. 
Writing  $\delta \Sigma^{\alpha}_{kl} =  \mathbb{S}^{(0)}_{kl}  [ e^{ i (\Omega t - \m k \cdot \hh x x_\alpha ) } -1]$
that is 
$\delta \Sigma^{\alpha}_{kl}  \approx i
 \mathbb{S}^{(0)}_{kl}     (\Omega t - \m k \cdot \hh x x_\alpha ) $
then 
$(\delta \Sigma^{\alpha}_{kl})^*  \approx - \delta \Sigma^{\alpha}_{kl}  $.
Similarly, 
we pose $\delta \Phi^{\alpha}  \approx i
 \rho^{(0)}     (\Omega t - \m k \cdot \hh x x_\alpha ) $.
  We obtain
as a result the net average speed
\begin{align}
& 
\langle
\dot{ C}_i
\rangle
\sim
\frac{\mathcal{G}^{(0)}_{jklu} }{24}  \times 
(\m k \cdot \hh x)
\int d  x_\alpha  \int d  x_\beta
 (
  x_\beta  -x_\alpha
)
U_{ijklu}
\label{eq:C-dot3}
 \end{align}
  where  
$ \mathcal{G}^{(0)}_{jklu} 
=[ 2  
 \mathbb{S}^{(0)}_{kl}  
\mathbb{S}^{(0)}_{uj} 
+  4 \rho^{(0)} \delta_{kl}  
\mathbb{S}^{(0)}_{uj}  ]
+ [ l  \leftrightarrow u].
 $
 The motion is along $\hh x$ here 
  but it could be more complex for other 
  choices of $\mathbb{S}^{(0)}_{kl}$. 
Thus the cell achieves motion by  propagating waves along the cell boundary. 
A wave breaks time
reversal symmetry \cite{Lau09}  and is consistent  with  the requirements for
swimming  at low Reynolds number \cite{Purcell}.

 \subsubsection{Numerical study}
 
  \begin{figure}[h!]
  \includegraphics[width=0.5\textwidth]{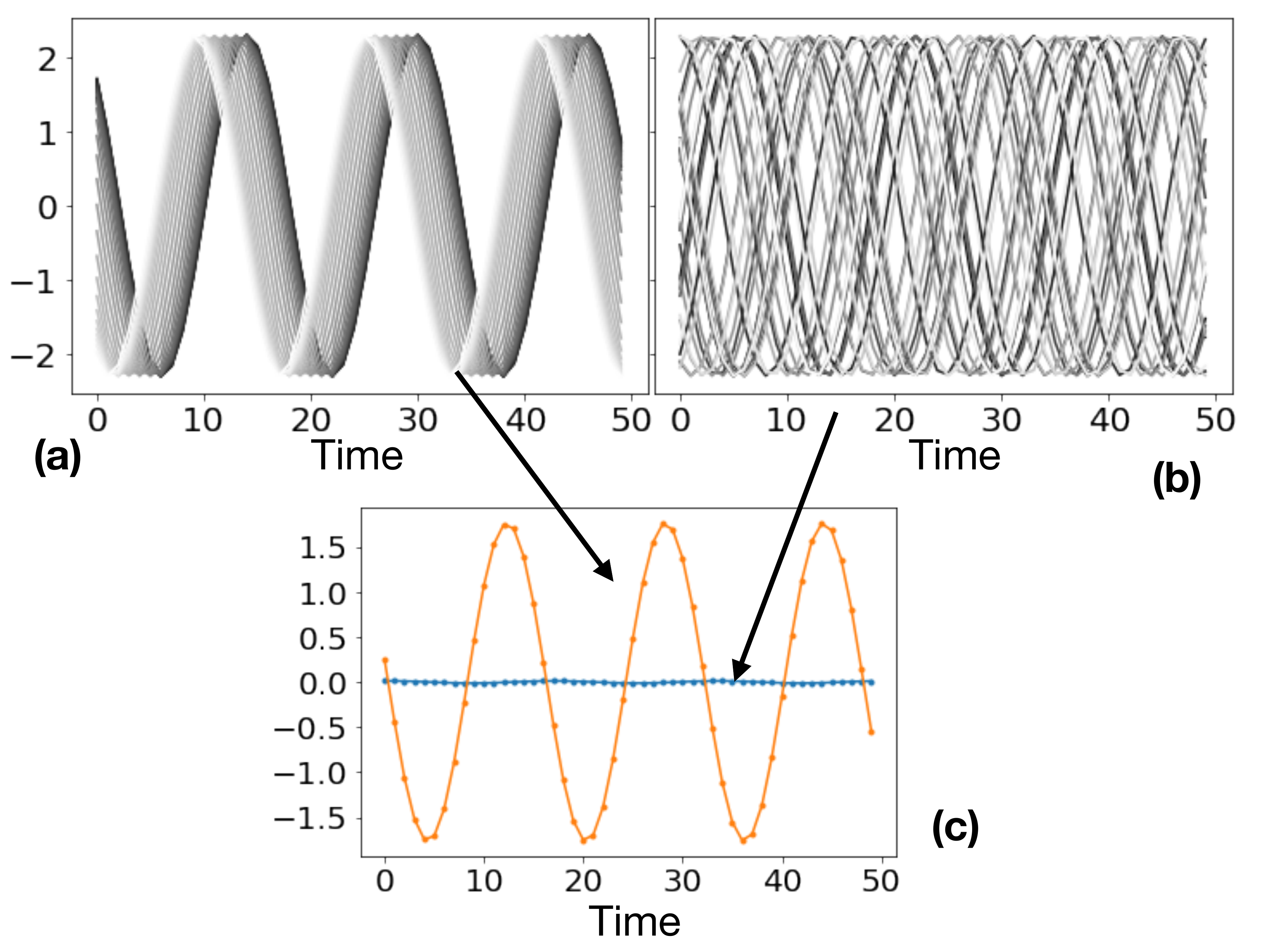}
  \caption{Comparison between in-phase  and anti-phase synchronization and their impact on the cell speed.
   (a) and (b) are amplitudes  oscillations  as function of time 
      for in-phase and anti-phase synchronization (obtained for $\chi  < 0$ i.e. $\mathcal{A}  < 0$  and $\chi >  0$ i.e. $\mathcal{A}  > 0$ in  \eq{eq:Fdot} respectively) considering the dynamics at long times. 
   The  oscillators lie at different positions along the $x$ axis. The different 
    oscillation amplitudes are shown on the same plot using different levels of grey.
       Amplitude and time
     are measured in 
   arbitrary units. 
  (c) Averaging all these amplitudes leads to collective oscillations, with low cell speed, for (a),  
  and no oscillations   with  higher cell speed -- see \fig{fig:numerics} -- for (b).   
  A similar phenomenology is observed in \cite{Weiner16} where the cell periphery was divided in 36 sectors, yielding 36 distinct actin oscillators. 
  Here we are considering $M = 33$ oscillators instead, reproducing approximately three periods 
  of oscillations. 
   }
   \label{fig:num1}
  \end{figure}
Results analogous to the conclusions of our model
 were reported  in an experimental study 
\cite{Weiner16}
  where cell motility was related to cellular oscillations and their synchronization,  see also \cite{Scita18}.
 In  \cite{Weiner16}, actin oscillators  distributed along the periphery of ameboid cells were observed. 
 By intervening on biochemical regulators,  the authors have been able to
  induce  strong coupling
  among the oscillators which  led to in-phase
   synchronization along the cell periphery.
  In this case,  as also expected from our analysis,
   migration is suppressed.
  For  normal untreated cells,  migration (and response to chemical gradients) was instead observed.

 \begin{figure}[h!]
\includegraphics[width=0.4\textwidth]{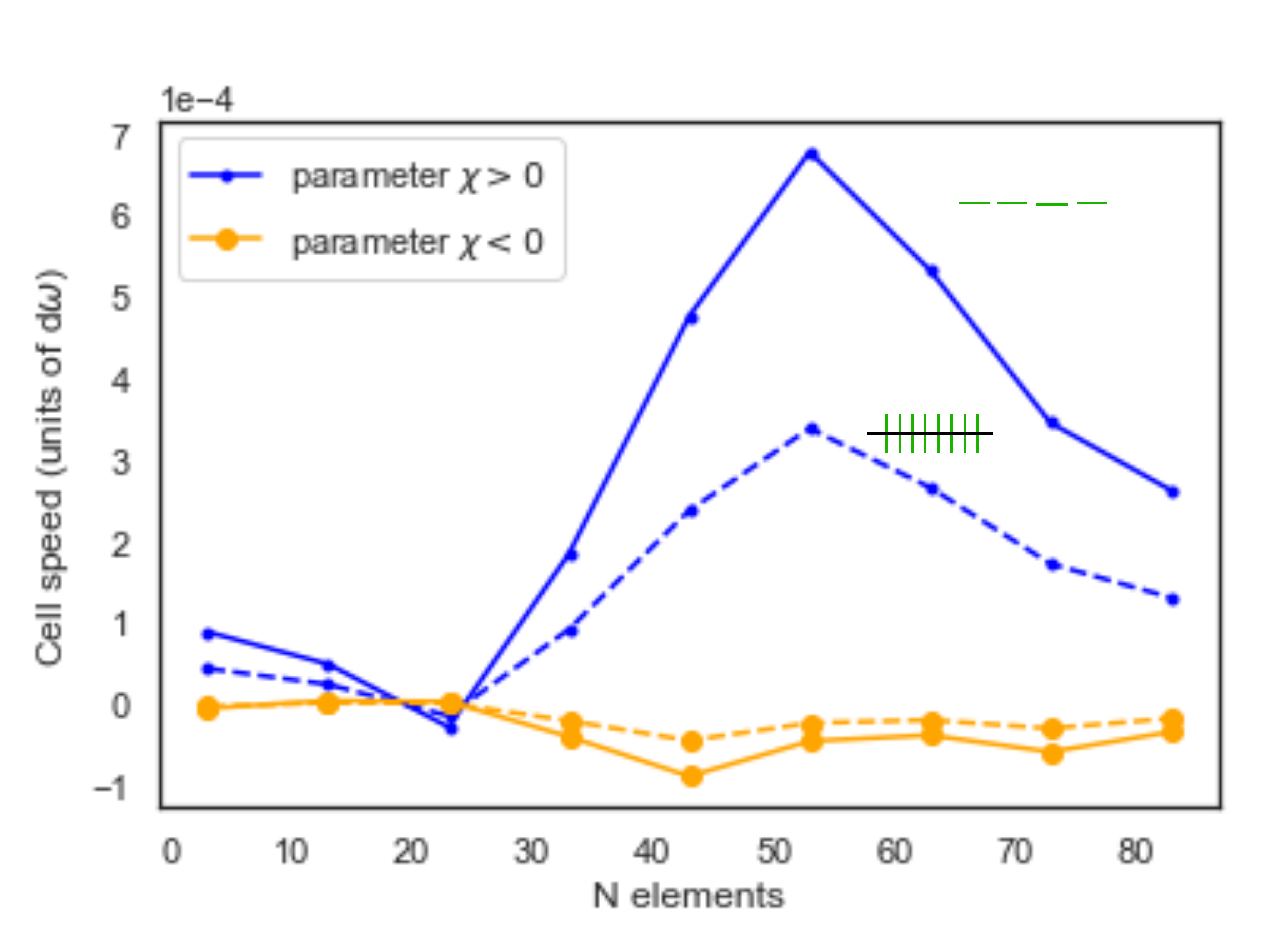}
\caption{Cell speed obtained from numerical simulations, investigating systems with 
different numbers of force generators. 
For simplicity 
all the orientations are held fixed along 
the same axis -- 
 either $\hh x$, continuous line,  or  $\hh y$, dashed line--  investigating two distinct scenarios : 
$\chi < 0$ (i.e. $\mathcal{A}  < 0$ in  \eq{eq:Fdot}) which here is expected to promote in-phase synchronization and 
$\chi > 0$ ( i.e. $\mathcal{A}  > 0$  in  \eq{eq:Fdot}) which here   is expected to promote anti-phase synchronization. 
The  cell speed is higher  in the case of anti-phase synchronization. 
Continuum (dashed) lines represent configurations where the orientations $\hh u_n$ are directed along the $\hh x$  ($\hh y$) axis.
}
\label{fig:numerics}
\end{figure} 

 In \fig{fig:num1} we show an example 
 (obtained from our numerics -- see appendix \ref{sec:app-cell-speed}  for details) 
 reproducing the phenomenology observed in experiments of  ref. \cite{Weiner16}:
  we consider $N=33$ 
   oscillators along a line  and study their collective behavior.
    In (a) oscillators are synchronized in-phase. 
  In (b) oscillators are synchronized in anti-phase.
We have computed  the cell speed 
 numerically 
 from \eq{eq:Cdot}, 
 and studied the speed 
 as a function of the non-isochrony parameter $\mathcal{A}  $ for different numbers of
  oscillators, i.e. different numbers  of  traction units.  
 We find  that when 
particles are synchronized in anti-phase,  $\mathcal{A}  > 0$ in \eqref{eq:Fdot},  
then the cell speed is higher than what observed in case of 
 in-phase synchronization, $\mathcal{A}  < 0$. 
 Although our analytical expression for the migration speed in
  \eqref{eq:Cdot2} indicates
  that the cell should not move  in case of   
   spatially homogeneous synchronization,
  the small motion observed in the numerics 
  might be due to inhomogeneities and instabilities of the ordered phase. 
  Instead,  anti-phase synchronization  
  leads to the propagation of waves \cite{LLPRE2012}, which 
 are  associated to a finite cell speed as in  \eq{eq:C-dot3}.
The reason  why  cell speed has a bi-phasic behavior as a function of the number of traction units, in \fig{fig:numerics},
 can be understood  using a scaling  argument reported in  appendix \ref{sec:app-cell-speed}.

\subsection{Dynamic equations for $\Phi$ and $\Sigma$ }
\label{sec:3}

In the previous section we have seen that the cell speed, the important readout for cell motility,
 depends on certain moments, defined in \eq{eq:mom-new}.  It becomes then important to discuss
 if, and under which conditions,
  these moments can attain finite values needed for motility. 
The growth of the moments 
 is  
related to a change of sign of a parameter in their dynamic equations which signals the  transition to order.
 Here we report  
  the dynamic equations for these moments  and
we find that 
  mechanical interactions, propagated via the substrate, among the traction units 
  can promote such a transition. 
  
The concentration $c$, in \eq{eq:c},  satisfies a Smoluchowski equation of the form of \eq{eq:Smol}
from which we derive 
 equations for the moments \eq{eq:rho} and \eq{eq:mom-new}, see appendix \ref{sec:app-moments} for the details.
 We find  that the relevant  moments 
 are $\Phi$ and $\Sigma$.  
 Density $\rho$ has a trivial equation due to our neglect of spatial gradients. 
We also find that the dynamics of $\bm{\Pi}$
shows no transition to order and is slave to that of $ \Sigma $. 
All this is expected as the microscopic model is made up of non-polar elements.

The dynamic equation for  $\Phi$
is 
\begin{align}
& \partial_t \Phi
\sim - D_\phi \Phi 
- \chi  \mathcal{Z}
\{
[ ( \frac{\delta_{il}}{2} \rho +  \mathbb{S}_{il} )
 \int dx_\alpha T_{ijkl}   \times 
\nonumber \\
&
( \frac{\delta_{jk}}{2} \Phi^{\alpha}  +   \Sigma^{\alpha}_{jk} ) ]
+
[k \leftrightarrow l ]
\}. \label{eq:Phi}
\end{align}
Here $  [ k \leftrightarrow l ] $ indicates the same contribution as in the previous parenthesis 
 but  exchanging $k$ and $l$.  $D_\phi$ describes phase diffusion. 
 Again  $\Phi^{\alpha}  = \Phi(x_{\alpha}, t) $ and
 we defined $ \mathcal{Z}  = \frac{ l^2_0 \omega_0 \gamma}{8 \beta} (1 + i \frac{ \beta}{\chi} ) $.
Note that in the aligned case, where $ \hh u^\alpha = \hh x$, 
the equation simplifies recovering
that of  oscillators in 1D \cite{LLPRE2012} although with some differences due to the  nature of the oscillators (dipolar here,
while in  \cite{LLPRE2012}  the oscillators were monopoles).

The dynamic equation for the  complex nematic tensor reads
\begin{align}
& \partial_t 
\Sigma_{fg} 
 \sim 
- 2 D \Sigma_{fg} 
 -\chi \frac{ \mathcal{Z} }{8}  \int dx_\alpha \Big( [  \{
\rho ( T_{fjkg}  + T_{gjkf}  )   \label{eq:Sigma}
 \\
& 
 + \frac{4}{3}   T_{ijkl} ( \delta_{fg} \mathbb{S}_{il} + perm_{fgil} )    
\}  (  \Phi^{\alpha} \frac{ \delta_{jk} }{2} + \Sigma^{\alpha}_{jk}  )  ]
+  [ k \leftrightarrow l ]  \Big)
\nonumber 
\end{align}
where $D$ describes rotational diffusion,  $perm_{fgil} = 
 \delta_{fi} \mathbb{S}_{gl} +  \delta_{fl} \mathbb{S}_{ig}  +   \delta_{ig} \mathbb{S}_{fl} +  \delta_{lg} \mathbb{S}_{if}  
$ and again $[ k \leftrightarrow l ]$ indicate the same contribution  as in the previous parenthesis but with 
  indexes $k$ and $l$ exchanged.

Both   \eq{eq:Sigma} and \eq{eq:Phi} depend on the nematic orientation tensor  $\mathbb{S}$. 
Its dynamic  equation, which is rather cumbersome and 
therefore reported in the appendix \ref{sec:app-S}, 
depends on $\Phi$ and $ \Sigma $.
Setting $ \partial_t 
\mathbb{S}_{fg} = 0 $ one can in principle obtain 
$\mathbb{S}$ as function of   
$\Phi$ and $ \Sigma $. 
That, in turn, allows one to obtain closed form equations for \eq{eq:Phi} and 
 \eq{eq:Sigma} which would describe the coupled dynamics of $\Phi$ and $\Sigma$.
 Instead of analyzing this general case, which involves lengthy and approximate calculations,
 we would like to summarize the generic aspects which 
 are relevant for the application of this 
 theory to the motility of cells.
 
 First,
considering  spatially homogeneous quantities, 
the term  $\rho^0 \Phi^{0} $  in \eq{eq:Phi}
is responsible for  the 
  transition
to order,  signaled by exponential growth of the order parameter $\Phi^0$. That
 is possible when the non-isochrony parameter
of \eq{eq:Fdot} satisfies $\mathcal{A} < 0$,
 i.e. $\chi <0 $.
However, as we discussed in  
\ref{sec:cell-speed-anal}, 
spatially homogeneous values of $\Phi$ are not enough for motility.
A finite cell speed, instead,
 is associated to wave-like perturbations, see \eq{eq:C-dot3}. 
These  could arise as 
 instabilities of the ordered state, as was discussed in \cite{ LLPRL2014}.
Alternatively, wave like behavior of the order parameter
 can be obtained in the opposite regime where $\mathcal{A} > 0$, i.e. $\chi > 0$, and the system's synchronization is  (close to)  
anti-phase 
 as our numerical study indicates, see \fig{fig:num1}.

Secondly,
 homogeneous order  in  \eq{eq:Sigma} 
 is possible 
thanks to the presence of a term  $\rho^0  \Sigma^{0}_{jk}  $.   
  The  transition to order (signaled by  exponential growth of the order parameter $\Sigma^0$)
  occurs 
   at some finite value of the density $\rho^0$ and requires
     $\chi < 0$, hence controlled by  
the sign of the non-isochrony parameter.
This is interesting as  it shows that 
the dynamic behavior  of   this novel complex order parameter $ \Sigma $ is to some extent related
 to that of $\Phi$. 
 Also here 
 a spatially homogeneous value of $\Sigma^0$ is not  enough for motility;
 wave-like perturbations, as examined in  \eq{eq:C-dot3},  might arise as instabilities of the ordered state \cite{ LLPRL2014}
or 
 be associated to other kinds of order (e.g. the analogue of anti-phase for this model).

\subsection{Discussion}
In this article we have developed  a theoretical formalism that relies on  
soft active matter and synchronization to
gain insight on the motility of cells. 
The theory combines vectorial degrees of freedom associated to the direction of the cell forces
and a phase variable associated to the time-dependent character of cell forces.
The force-free constraint, saying  that cellular forces  add up to zero, 
led us to introduce a novel broken symmetry variable -  a complex tensor field -
to  model  the distribution of cellular  traction force.

 We computed the  cell speed and found that it also
  depends on this tensor field as well as on the
   complex order parameter
    describing synchronization
  in Kuramoto model.
 We found that  
 in-phase synchronization is not enough
to promote motility.  
    Our model requires more complex patterns,  
     e.g. the propagation of waves. 
    These might arise from
     1)  hydrodynamic instabilities of the ordered synchronized state
    or  2)  instabilities  of the anti-phase state. 
     We have performed  numerical analysis which confirms this view and
      suggests that both mechanisms are possible, with 2) giving higher speed. 
      
 To make further progress, and test our predictions regarding the correlation between 
 cell speed and its relation with $\Sigma$ and $\Phi$,
    it would be helpful to compute 
      $\Sigma$ 
    directly from experimental traction patterns of migrating cells. 
   Note instead that for the case of adherent cells, which do not migrate, the complex tensor description  is a priori not necessary.

Here we have provided a realization of cell with subcellular traction units that implement, locally, the force-free condition. 
This allows us to simplify the analysis and to derive general analytical expressions. Other types of force-distributions, where the
 force-free condition is not satisfied locally, but only recovered  at the cell scale,
  are expected to give qualitatively similar
  results,  with quantitative differences in \eq{eq:Phi} and  \eq{eq:Sigma}.

   Finally,  
  here  we have modeled the extra-cellular environment as 
   a fluid substrate.
  A more realistic description requires to model
   an elastic, adhesive, substrate. 
  Migration on such  complex environments can be tackled
     relying on the  formalism we have developed here,  by  modifying the 
 force balance, \eq{eq:force-balance}, to include  the adhesion's dynamics \cite{LS17}
and by extending that approach  to  elastic substrates.
This  task is left for a future work.

\acknowledgments{The author acknowledges 
financial support from the ICAM Branch Contributions and Labex Celtisphybio No. ANR-10- LBX-0038 part of the IDEX PSL No. ANR-10-IDEX-0001- 02 PSL 
and many stimulating scientific discussions with T. B. Liverpool, 
 M. C. Marchetti, and P. Sens.
}

\section*{APPENDIX}

In this section 
 we present some details which complement  the main text.

\subsection{Dynamics of the force generations}

 \paragraph{Orientational dynamics}
\label{sec:app-orient}
In \eq{eq:omega}
\begin{align}
&   \mathcal{O}^{\alpha \beta}_i  \approx - 
 \frac{R_0  f_0 l^2_0}{192 }  \sin(\phi_\alpha - \phi_\beta)
 \ha  u^\beta_j
[\ha  u^\alpha_k  \ha  u^\alpha_l  \ha  u^\alpha_u  \ha  u^\beta_v+  
 \label{eq:O}      \\
&
+ \ha  u^\alpha_k  \ha  u^\alpha_l  \ha  u^\alpha_v  \ha  u^\beta_u 
+ \ha  u^\alpha_k  \ha  u^\alpha_u  \ha  u^\alpha_v  \ha  u^\beta_l 
+ \ha  u^\alpha_l  \ha  u^\alpha_u  \ha  u^\alpha_v  \ha  u^\beta_k
   ]
V_{ij kl uv}(x_\alpha-x_\beta )
\nonumber 
 \end{align}
where  
$
V_{ij kl u v}(x_\alpha-x_\beta )
= 
\nabla_v \nabla_u  \nabla_k \nabla_l H_{ij}[(x_\alpha-x_\beta )\hh x]
.  $

\paragraph{Traction unit deformation dynamics}
\label{sec:app-def}

The dynamic equation for the  internal deformation of the element $\alpha$ is given by  \eq{eq:Ldot}.
There, $ \bm{ \mathcal{I}}^{\alpha \beta}    $ follows 
 from force balance, \eq{eq:force-balance}, and 
to  the leading order is given by  
\begin{equation}
  \mathcal{I}^{\alpha \beta}_i  \approx - 
 L^\alpha  L^\beta
(\ha  u^\alpha_k  \ha  u^\beta_l +  \ha  u^\alpha_l  \ha  u^\beta_k   ) 
      \ha u^\beta_j
     F^\beta  
     T_{ij kl}(x_\alpha-x_\beta )
 \label{eq:I}    
 \end{equation}
where  the tensorial part is 
$
 T_{ij kl} (x_\alpha-x_\beta )
= 
 \nabla_k \nabla_l H_{ij}[(x_\alpha-x_\beta )\hh x]
 $ as above in the main text.

\paragraph{Amplitude and phase dynamics}
\label{sec:app-phase-ampl}
The equations for amplitude and phase
   $\dot{R}_\alpha$
and  
  $\dot{\phi}_\alpha$, 
  describing the time-dependent oscillatory dynamics
 are  obtained
 by 
mapping 
$L^\alpha$ and $F^\alpha$ onto amplitudes  ${R}_\alpha$ and phases     $\dot{\phi}_\alpha$.
 This is achieved by introducing complex amplitudes  $A^\alpha $ related to the deformations and forces via
  $ 
  L^\alpha \sim l_0 +  [ A_\alpha e^{i \omega t}  + c.c. ]/2$
and
$F^\alpha \sim i  \gamma \omega_0 [ A_\alpha e^{i \omega t}  - c.c. ]/4$,  
 as explained in the main text. Thanks to this change of variable, we
 obtain dynamic equations for   $ \dot{A}_\alpha $  and $ (\dot{A}_\alpha )^*$.
 Finally, writing 
 $A_\alpha = R_\alpha e^{ i \phi_\alpha} $ we derive equations for the amplitudes
 from 
  $ \dot{R}_\alpha  = 
  \frac{ e^{- i \omega t}   \dot{A}_\alpha + 
  e^{ i \omega t}   (\dot{A}_\alpha )^*
  }{2  }
  $
and  for the phases from 
 $\dot{\phi}_\alpha  =
  \frac{ e^{- i \omega t}   \dot{A}_\alpha -
  e^{ i \omega t}   (\dot{A}_\alpha )^*
  }{2 i R_\alpha }
  $.

Setting $ \dot{ \delta R}_\alpha = 0 $ in \eq{eq:Rdot}  
we can eliminate  $ \delta R_\alpha $
 in favor of the phase  variable obtaining
 $$ \delta R_\alpha \sim 
 l^2_0    \ha u^\alpha_i \ha u^\beta_j (  \ha u^\alpha_k \ha u^\beta_l +  \ha u^\alpha_l \ha u^\beta_k  )
\frac{R_0 \omega_0 \gamma}{8 \lambda } \sin( \phi_\beta  - \phi_\alpha )
T_{ijkl }\nonumber    $$
Using this expression, we get
\eq{eq:Phidot2}

\subsection{Cell Speed}

\label{sec:app-cell-speed}

\subsubsection{Analytical study}

In the equation \eq{eq:Cdot}: 
to leading order we find
$ 
  \dot{C}  \propto 
\frac{1}{M}
\sum^M_{\alpha=1} \sum_{\beta \neq \alpha} 
 \nabla^3(\frac{1}{r_{\alpha \beta}}) L^\alpha   \dot{d}^\beta.
$
The term
$\nabla^3(\frac{1}{r_{\alpha \beta}})  $
scales as 
$\nabla^3(\frac{1}{r_{\alpha \beta}}) \sim \frac{1}{r^3_0(\alpha -\beta)^3} $
where $r_0$ is the average separation between the centres of the traction units (here chosen to be constant),
while 
the term  $L^\beta   \dot{d}^\alpha $
scales as
$L^\alpha   \dot{d}^\beta \sim  \omega_0 d^2_0 \sin  (\beta -\alpha)  $ where
 $d_0$ and $\omega_0$ are the oscillation amplitude and frequency.
 The sum over the index $\alpha$ brings $M$ 
such contributions which cancel
 the $1/M$ pre-factor. 
To study the cell speed as a
 function of $M$ we can estimate 
the previous expression as 
$$ 
| \dot{C} |
 \propto 
| 
\int^M_1 dx \frac{ \sin x }{x^3}  
|
\sim 
[ \frac{ \sin x }{2x^2}  
+ \frac{ \cos x }{2x}
+ \frac{1}{2} Si(x)]^M_1 . 
$$
For small values of  $M, M \sim 1 + \epsilon $ the speed $| \dot{C} |$ grows linearly with 
$\epsilon$. For large values of 
 $M$, $| \dot{C} |$ saturates as   
$Si(M) \to \pi$ while 
the remaining terms in parenthesis tend to zero. Taken together these limits
 provide an explanation for the bi-phasic behavior seen in \fig{fig:numerics}.

\subsubsection{Numerical study}
We  integrate 
 \eq{eq:Ldot},  \eq{eq:Fdot},  \eq{eq:Cdot} using Euler scheme with time-step $dt = 0.0031 s$.
Parameters characterizing the dynamics of individual
 oscillators are: $\mathcal{K}  = 1 pN /(\mu m  \quad s)$,  
 $ \mathcal{S} = 1/(\mu m)^2  $, 
  $\mathcal{M} = 0.01 pN /\mu m  $ and $\mathcal{A} = \pm 0.01  pN /(\mu m^3  \quad s) $.
The drag coefficient is $\gamma = 3 \pi \eta a$ as in \cite{LS15} (ignoring for simplicity cell viscosity) 
where $a $ is the radius of each adhesion site,
$a = 0.1 \mu m$. The viscosity of the substrate is chosen to be $\eta \sim  6.3 \eta_{H_2 O}$ where
  $ \eta_{H_2 O}$ is water's viscosity and  
  interactions are modelled as $ H (r) = \frac{1}{2  \pi \eta r}$,  \cite{LS15} , 
with  average separations among units  
 $ r_0 = 25 \mu $m and equilibrium length $ l_0 = 15 \mu $m.

\subsection{Dynamic equations for the moments}

\label{sec:app-moments}
The concentration $c$, in \eq{eq:c}, satisfies a Smoluchowski equation of the form \cite{Marchetti-Rev-Mod-Phys}
\begin{align}
& \partial_t c =
 \bm{\mathcal{R}} \cdot 
[ 
D 
\bm{\mathcal{R}} 
 c
- 
c \bm{\Omega}
] 
+ 
\frac{\partial}{ \partial \phi } 
[ 
D_\phi \frac{\partial}{ \partial \phi}  c
- 
c \Psi
]
\label{eq:Smol}
\end{align}
where $D$ and $D_\phi$ are respectively
 the rotational and the phase 
diffusion constants.
Note 
that terms of the form $ \bm{\nabla} \cdot [ v_0  \hh u   c ]$, 
which are typical non-equilibrium terms describing self-propelling particles (with speed $v_0  \hh u $ ),
are absent here as the individual   elements do not self-propel.

The two terms in parenthesis $[ ... ]$ at r.h.s. are respectively
  rotational and phase currents. Here $ \bm{\mathcal{R}}  $ is defined as 
  $ \bm{\mathcal{R}}  := \hh u \wedge \frac{\partial }{\partial  \hh u }$, while 
 $ D_\phi \frac{\partial}{ \partial \phi}  $ describes phase diffusion.
The other terms in parenthesis  
comprise
\begin{align}
&  \bm{\Omega} =
\int d  x' \int d\hh u' 
\int d\phi' 
\bm{\omega}
 c(\m x', \hh u',\phi', t) ]
\end{align}
and
\begin{align}
& \Psi =
\int d x' \int d\hh u'  \int d\phi'  {\psi}  c(\m x', \hh u',\phi', t)
\end{align}
where 
$\bm{\omega}$
and  $\psi$ 
are  given by \eq{eq:omega} and \eq{eq:Phidot2}.

For the angular dynamics, we use the expression of 
 \eq{eq:omega}
and get 
\begin{align}
& \Omega_a(x)  \sim  
\epsilon_{a b i}
\ha u_b
\frac{R_0 f_0 l^2_0}{384 i} 
\int dx_\beta 
V_{ij kluv}(x -x_\beta) \{
 [ (\frac{\delta_{jv} }{2}  \Phi^\beta  + 
\nonumber \\
 & 
    \Sigma^\beta_{jv} )   \ha u_k  \ha u_l   \ha u_u e^{-i \phi } 
     -c.c.
    ]
    + [v \leftrightarrow k ]
     + [v \leftrightarrow l ]
       + [v \leftrightarrow u ] 
      \} \nonumber 
\end{align}
where 
$c.c.$ denotes the complex conjugate and 
 $ [v \leftrightarrow k ]$
 indicates the same contribution as the one in the previous parenthesis but with 
 $v$ replaced by $k$ etc. 
   We find that $\Omega_a $  is
  sub-dominant (in terms of powers of the inverse separation which controls the 
 interaction strength) in the equations
for the moments %$\m P$,
 $\bm{\Pi}$ and $\Sigma$.
In fact the dominant contribution comes from the 
term $\Psi$ associated to the 
 term describing the two-body phase dynamics. 

The phase dynamics
is obtained from \eq{eq:Phidot2}
as 
\begin{align}
& \Psi(x)
\sim 
-\frac{\chi  l^2_0 \omega_0 \gamma }{8 \beta i}  \int dx_\beta  
T_{ijkl} (x- x_\beta )  \times
\nonumber \\
& 
\{ 
[ \ha u_i \ha u_l 
(\frac{\delta_{jk} }{2} 
\Phi^{\beta}
+
\Sigma^{\beta}_{jk} ) e^{-i \phi } - c.c.  ]
+ [k \leftrightarrow l]
\}
\end{align}

Below we present the dynamic equations for the moments, \eq{eq:rho}, \eq{eq:mom-new}.
For example, the equation for the moment $\Phi$
is obtained from
$ \dot{\bm{\Phi}}  =  
\int d x 
  e^{i\phi}
\partial_t c  
$ and by inserting, in this expression, the r.h.s. of \eq{eq:Smol}.
 A similar procedure is followed for the other moments.

\subsubsection{Equation for $\rho$}
Neglecting, as we did,
 terms containing the spatial gradients
then the density equation is simply
$\partial_t \rho = 0.$

\subsubsection{Equation for $p$}

The dynamic equation   for $\m p$ 
 is
\begin{align}
& 
\partial_t  p_a =  - D  p_a
+ 
\frac{l^2_0 f_0}{384 i }
(\delta_{bp} \delta_{ia} - \delta_{ba} \delta_{ip}   ) \times
\nonumber \\
&
  \{ 
\Gamma_{bklupf}  \frac{\Pi^*_f}{24} 
 \int dx_\beta 
V_{ijkluv} (x_\alpha  -x_\beta )
 [  \delta_{jv}  \Phi^\beta +  \Sigma^\beta_{jv}   ]
- c.c.
 \} \nonumber 
\end{align}
where
%and 
$ \Gamma_{bklupf}   = 
\delta_{bk } \Delta_{lupf} + 
\delta_{bl } \Delta_{kupf} 
+ \delta_{bu} \Delta_{lkpf} 
+  \delta_{bp} \Delta_{lukf} 
+ \delta_{bf} \Delta_{lupk}
$
and
$\Delta_{abcd} = \delta_{ab}  \delta_{cd}  +  \delta_{ac}  \delta_{bd} +  \delta_{ad}  \delta_{bc} $.
Hence,  the dynamics of   $\m p$ is  slave
to that of other moments. 
Moreover,  when the particles are synchronized, the interaction terms vanish. 
This is  expected based on the symmetry of the microscopic elements which are nematic
rather than polar.  
One can see that by considering that the quantity $\bm{\Pi}$, in the synchronized case,
 reads $\bm{\Pi} = \Phi^0 \m p $. Hence  $\bm{\Pi}^* = (\Phi^0)^* \m p $ 
and terms of the form $ \bm{\Pi}^* \Phi^0 - c.c. $ vanish in that case.
A similar consideration holds true also  for terms of the form $ \bm{\Pi}^* \cdot   \Sigma $.

\subsubsection{Equation for $\Pi$ }

The equation   for $\bm{\Pi}$  is 
obtained from 
$ \dot{ {\Pi}}_f  =  
\int d x 
 \ha u_f e^{i\phi}
\partial_t c  
$. 
To get this equation we use the
 r.h.s. of \eq{eq:Smol} neglecting the term  
 $ c \bm{\Omega} $ which is subdominant (as powers of the inverse of the distance, which characterizes the
  coupling strength)  compared to $c \Psi $.
We obtain
\begin{align}
& \partial_t  \Pi_f = 
 - D  \Pi_f  -\frac{\mathcal{Z}}{4} \times
  \label{eq:Pi} \\
 &
 \int dx_\alpha \{ [ T_{ijkl} (x_\alpha - x) \Delta_{ijkf} p_f  ( \frac{\delta_{jk}}{2} \Phi^\alpha +  \Sigma^\alpha_{jk}  ) ]+
 [  l \leftrightarrow k ]
  \} \nonumber 
\end{align}
Note that also here the dynamics of $\bm{\Pi}$ is slave to the dynamics of $ \Sigma $ and $\Phi$  and 
no transition 
to spatially homogeneous  order is obtained. 
All this is expected as the microscopic force generators are non-polar elements.

\subsubsection{Equation for $ \mathbb{S} $ }
\label{sec:app-S}

The dynamic equation for $\mathbb{S}$ is given by
\begin{align}
& \partial_t 
\mathbb{S}_{fg}
 \sim 
 \{
-  D \mathbb{S}_{fg} 
\nonumber \\
&
 + \frac{f_0 l^2_0}{384 i}
(\delta_{bp} \delta_{if} - \delta_{bf} \delta_{ip}  ) 
  \int dx_\alpha
 V_{ijkluv} 
\Big( 
  [\frac{\delta_{jv} }{2} \Phi^\alpha +    \Sigma^\alpha_{jv}  ]
\nonumber \\
& 
\times 
[\frac{ \Gamma_{pgbklu} }{48}
 \Phi^*
+ \frac{ 1 }{48} (  \Sigma^*_{pg}  \Delta_{bklu} + perm_{pg bklu } )
 ]
 -c.c.
\Big) \} 
\nonumber 
\\
& + 
\{ f \leftrightarrow g \}
\label{eq:S}
\end{align}
which
 shows that the dynamics of $\mathbb{S}$ is slave to that of $\Sigma$ and $\Phi$.
 Here the term $perm_{pg bklu } = 
\Sigma^*_{bg}  \Delta_{pklu}  + 
\Sigma^*_{kg}  \Delta_{bplu} 
+
\Sigma^*_{lg}  \Delta_{bkpu}
+
\Sigma^*_{ug}  \Delta_{bklp} .$ 
Again, $ \{ f \leftrightarrow g \}$ represents the same contribution of
 the  terms in the previous parenthesis with indexes $f$ and $g$ exchanged.

\bibliography{notes}

\end{document}